\documentclass{eas}
\usepackage{graphicx}
\def\gs{\mathrel{\raise0.35ex\hbox{$\scriptstyle >$}\kern-0.6em \lower0.40ex\hbox{{$\scriptstyle \sim$}}}}
\def\ls{\mathrel{\raise0.35ex\hbox{$\scriptstyle <$}\kern-0.6em \lower0.40ex\hbox{{$\scriptstyle \sim$}}}}
\newcommand{\Msolar}{\mbox{$\rm M_{\odot}\,$}}
\newcommand{\Lsolar}{\mbox{$\rm L_{\odot}\,$}}
\newcommand{\arcsecs}{\mbox{$^{\prime\prime}$}}
\begin{document}

\title{The ISM in distant galaxies} 
\author{Thomas R.\ Greve}\address{Max-Planck Institute f\"ur Astronomie, K\"onigstuhl 17, D-69117 Heidelberg, Germany}
%
%
\begin{abstract}
The interstellar medium (ISM) is a key ingredient in galaxy formation and
evolution as it provides the molecular gas reservoir which fuels star formation
and supermassive black hole accretion. Yet the ISM is one of the least studied
aspects of distant galaxies.  Molecular and atomic transitions at
(sub)millimetre wavelengths hold great promise in measuring macroscopic
properties (e.g.\ masses, morphologies, star formation laws), as well as
microscopic properties (e.g.\ gas densities, temperatures, cooling) of high-$z$
galaxies.  In this overview I summarize the growing number of high-$z$
molecular line detections, highlighting some of the most intriguing results
along the way. I end by discussing a few areas where future facilities (e.g.\
ALMA, EVLA, CCAT, LMT) will drastically improve on the current state of affairs.
\end{abstract}
\maketitle
\section{Introduction}
In the 17 years since the first discovery of $^{12}$CO (hereafter CO) at high
$z$ in IRAS\,F10214$+$4724 (Brown \& Vanden Bout 1992; Solomon, Radford \&
Downes 1992), a serendipitously discovered type-2 QSO at $z=2.29$, little more
than 60 molecular line detections at $z>1$ have been made (Fig.\ 1). This
relatively modest progress is in spite of early search campaigns (Evans et al.\
1996; van Oijk et al.\ 1997), and reflects an unfortunate combination of
limited receiver sensitivity and the lack of atmospheric transmission in
certain (sub)mm bands, which makes such observations extremely challenging.
For the same reasons early efforts were primarily limited to CO observations of
extremely luminous, optically selected QSOs (e.g.\ Ohta et al.\ 1996; Omont et
al.\ 1996) and high-$z$ radio galaxies (HzRGs -- e.g.\ Papadopoulos et al.\
2000; De Breuck et al.\ 2003).  In order to enhance the chances of success, the
targeted objects were often strongly lensed systems and/or pre-selected to have
strong thermal dust emission.

These first molecular line detections in QSOs and HzRGs were of great
importance as they unambiguously demonstrated the existence of large molecular
gas reservoirs enriched in C and O in the early Universe.  A prime example is
the detection of CO in the highest known redshift QSO (J114816.64$+$525150.3 at
$z=6.419$ -- Walter et al.\ 2003; Bertoldi et al.\ 2003), which implies
significant enrichment of the gas at a time when the Universe was only 1/16th
of its present age and had only recently been reionized.  Many of these systems
have since been revisited in more detail, resulting in remarkable new results
such as exquisite high-resolution CO images (e.g.\ Carilli et al.\ 2002;
Riechers et al.\ 2008), nearly fully sampled CO spectral line distributions
(SLEDs) (Wei\ss~et al.\ 2007), and in some cases detections of 
HCN and HCO$^+$ (e.g.\ Solomon et al.\ 2003; Riechers et
al.\ 2006) as well as atomic transitions of C\,{\sc i} and C\,{\sc
ii} (e.g.\ Barvainis et al.\ 1997; Maiolino et al.\ 2005).

Parallel to these efforts, CO searches were carried out in high-$z$ galaxies
dominated by star formation, resulting in the first detection of molecular gas
in submm-selected galaxies (SMGs) (Frayer et al.\ 1998, 1999). The advent of a
large sample of SMGs with spectroscopic redshifts (Chapman et al.\ 2005)
allowed for the first systematic CO survey of a well defined sample of high-$z$
galaxies (Neri et al.\ 2003; Greve et al.\ 2005; Tacconi et al.\ 2006, 2008).
The result was robust estimates of the typical gas masses, sizes, kinematics
and gas depletion time scales for bright SMGs. CO observations have also been
carried out of three optically-faint radio galaxies (OFRGs) -- thought to be
'hot-dust' versions of SMGs as they are selected to be luminous in the radio
but faint at 850-$\mu$m wavelengths -- and towards two sBzK galaxies, which are
massive, but moderately star forming, disk-like galaxies at $z>1$ (Daddi et
al.\ 2008).
\begin{figure}
\includegraphics[angle=0,width=1.0\hsize]{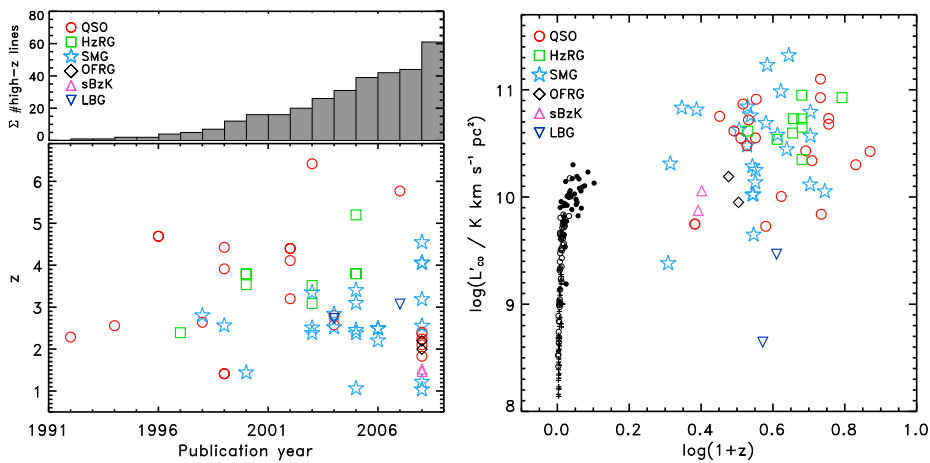}
\caption{{\bf Left:} The cumulative distribution of the number of $z>1$
molecular line detections in the past 17 years (top panel), and their galaxy
type distribution in chronological order (bottom panel). {\bf Right:} CO
luminosities vs.\ redshift for $z>1$ galaxies and for a comparison sample of
local LIRGs and ULIRGs (Sanders et al.\ 1991; Solomon et al.\ 1997; Yao et al.\
2003). Objects have been corrected for lensing where appropriate.}
\end{figure}
Finally, the recent detections of CO in a few rare examples of strongly lensed
Lyman Break Galaxies (LBGs) (Baker et al.\ 2004; Coppin et al.\ 2007) have 
allowed for a unique peak into the gas
properties of 'ordinary' high-$z$ galaxies, and the kind of studies which will
be possible with ALMA.

\section{The molecular gas content of distant galaxies}
Molecular hydrogen (H$_2$) is by far the main component of molecular clouds and
the molecular ISM in general. The lack of permanent dipole moment of the H$_2$
molecule, however, allows only quadrupole $\Delta J = 2$ transitions, with
$S(0): J_u - J_l = 2 - 0$ at 28\,$\mu$m being the easiest to excite, but with
its $\Delta E_{20}/k_B = 510\,$K much too high for the gas temperatures
prevailing in the bulk of the molecular gas ($\sim 15-60\,$K). Moreover, the
Earth's atmosphere is largely opaque at 28\,$\mu$m. Consequently, the next most
abundant molecule, CO with a Galactic abundance of $[\mbox{CO}/\mbox{H}_2] \sim
10^{-4}$, is used. Its permanent dipole moment, and easily excitable rotational
transitions (e.g.\ for $J=1-0$: $\Delta E_{10}/k_B\sim 5.5\,$K) with
frequencies at mm and submm wavelengths where neither dust in the Galaxy, nor
the Earth's atmosphere are significantly absorbing, make CO the ideal bulk
H$_2$ mass tracer.

The high optical depths of the CO transitions do not prevent them from tracing
the entire volume and mass of an ensemble of clouds where large velocity
gradients keep $\Delta v_{\mbox{\tiny{cloud}}} >> \Delta
v_{\mbox{\tiny{thermal}}}$, and thereby the entire ensemble fully transparent
to CO line emission (i.e.\ the high optical depths arise only locally). If the
additional assumption of virialized clouds is made then the total H$_2$ gas
mass ($M_{\mbox{\tiny gas}}$) is directly proportional to the total CO
luminosity ($L'_{\mbox{\tiny CO}}$): $M_{\mbox{\tiny gas}} = \alpha
L'_{\mbox{\tiny CO}}$ (e.g.\ Dickman et al.\ 1986), where $\alpha$ is the
CO-to-H$_2$ conversion factor and depends on the CO brightness temperature
($T_{\mbox{\tiny{b}}}$) and the gas density ($n$) according to $\alpha \propto
\sqrt{n}/T_{\mbox{\tiny{b}}}$.  Three decades of Galactic CO observations have
established a CO-to-H$_2$ conversion factor that applies to molecular clouds in
our Galaxy ($\alpha_{\mbox{\tiny{G}}} =
4.6\,\Msolar/(\rm{K\,km\,s^{-1}\,pc^{2}})$ -- this includes a 36\% correction
for He). The fact that denser clouds are usually warmer keeps
$\alpha_{\mbox{\tiny{G}}}$ constant within a factor of $\sim 2$ (e.g.\ Young \&
Scoville 1991).  However, in local Ultra Luminous Infra-Red Galaxies (ULIRGs),
which harbour extreme tidal fields and radiation pressures capable of disrupting
molecular clouds, the  appropriate conversion factor is
$\alpha_{\mbox{\tiny{IR}}} = 0.8\,\Msolar/(\rm{K\,km\,s^{-1}\,pc^{2}})$, i.e.\
$5\times$ smaller than the Galactic value (Solomon et al.\ 1997).

What CO-to-H$_2$ conversion factor should be used to infer the gas masses of
high-$z$ galaxies? It is usually assumed that conditions at high redshifts are
akin to those in local ULIRGs, and high-$z$ gas masses are therefore derived
using $\alpha_{\mbox{\tiny{IR}}}$. However, $\alpha_{\mbox{\tiny{IR}}}$ was
deduced from a small sample, and particular relations (such as
$M_{\mbox{\tiny{new stars}}}=2/3 L'_{\mbox{\tiny{CO}}}$) were assumed to derive
it (Downes \& Solomon 1998) -- relations which may not apply within the wider
local ULIRG class, let alone their high-$z$ counterparts.  Even if
$\alpha_{\mbox{\tiny{IR}}}$ is appropriate for extreme, IR-luminous galaxies,
which make up the bulk of the high-$z$ CO detections, a conversion factor
closer to $\alpha_{\mbox{\tiny{G}}}$ probably apply to the recent CO detections
towards LBGs and sBzK galaxies. In the metal-poor H$_2$ gas expected in LBGs,
for example, CO may photodissociate while leaving the largely self-shielding
H$_2$ clouds intact.  In fact, a high-resolution CO study of a combined sample
of SMGs, sBzK/BX and LBGs at $z\sim 2-3$ found that only by assigning a
Galactic conversion factor to the UV/optical selected galaxies and a ULIRG
conversion factor to the SMGs, could the dynamical masses be reconciled with
realistic gas fractions (Tacconi et al.\ 2008).

\smallskip

The published high-$z$ CO detections to date yield median CO luminosities
of QSOs, HzRGs, and SMGs of $\langle L'_{\mbox{\tiny CO}}\rangle = 3.5\times
10^{10}$, $4.5\times 10^{10}$, and $3.7\times
10^{10}\,\mbox{K}\,\mbox{km}\,\mbox{s}^{-1}\,\mbox{pc}^2$, respectively.
Assuming $\alpha = \alpha_{\mbox{\tiny{IR}}}$ and thermalized transitions,
meaning CO line ratios of unity, the corresponding gas masses become $\langle
M_{\mbox{\tiny{gas}}}\rangle=2.8\times 10^{10}$, $3.6\times 10^{10}$, and
$3.0\times 10^{10}\,\Msolar$.  The median CO luminosities found for OFRGs,
sBzK, and LBG galaxies are $\langle L'_{\mbox{\tiny CO}}\rangle = 1.6\times
10^{10}$, $1.1\times 10^{10}$, and $2.9\times
10^{9}\,\mbox{K}\,\mbox{km}\,\mbox{s}^{-1}\,\mbox{pc}^2$, respectively, with
corresponding gas masses of $\langle M_{\mbox{\tiny{gas}}}\rangle=1.3\times
10^{10}$, $8.8\times 10^{9}$, and $2.3\times 10^{9}\,\Msolar$.  The bulk of
$z>1$ galaxies with CO detections have $L'_{\mbox{\tiny CO}} > 1\times
10^{10}\,\mbox{K}\,\mbox{km}\,\mbox{s}^{-1}\,\mbox{pc}^2$, which is $\sim
3-5\times$ larger than local ULIRGs (Solomon et al.\ 1997) and reflects the
bias towards luminous systems at high redshifts.  

Currently, a direct comparison of the gas masses in local and distant galaxies
is complicated by fact that while local galaxies are readily detected in the
low CO lines ($J = 1-0, 2-1$), the high-$J$ lines are difficult to access due
to the atmosphere. For distant galaxies, the situation is reverse: most
high-$z$ galaxies are first detected in high-$J$ lines ($J = 3-2$ or higher),
which are the brightest lines and are redshifted into clean atmospheric
windows. Only then are searches for CO $J=1-0$ attempted.  All low-$J$
detections to date have been of sources with previous detections in higher CO
lines. This introduces a bias in the sense that at high redshifts, galaxies
with dense and warm gas are preferred over those with a more quiescent ISM.
Furthermore, it means that the CO $J=1-0$ luminosity, and therefore the gas
mass, must be inferred from assumptions about the excitation conditions of the
gas.  For example, the gas masses inferred from CO $J=1-0$ observations towards
the two distant starbursts ERO\,J16450$+$4626 ($z=1.44$) and SMM\,J13120$+$4242
($z=3.41$) (Greve, Ivison \& Papadopoulos 2003; Hainline et al.\ 2006) were
substantially larger ($4-10\times$) than previous estimates based on CO $J=5-4$
and the assumption that the line was thermalized.  Similar claims has been made
for a few QSOs (Papadopoulos et al.\ 2001), although more recent observations
find no evidence for cold gas reservoirs in QSOs (Riechers et al.\ 2007).

\section{Spatially and kinematically resolved ISM observations at high $z$}
Spatially and kinematically resolved molecular line observations hold enormous
potential for direct imaging of galaxy assembly in the distant Universe -- a
process for which several distinct scenarios can be envisaged.  Does it occur
via numerous discrete short bursts within a huge gravitationally bound gas
reservoir, or via multiple mergers of gas-rich galaxies ('wet' mergers)
interspersed with prolonged quiescent intervals, or perhaps a single widespread
starburst ('monolithic' collapse)?  Resolved gas-kinematics also provide
estimates of galaxy sizes, dynamical masses, merger fraction etc., key
quantities to test against model predictions.

\begin{figure}
\includegraphics[angle=0,width=1.0\hsize]{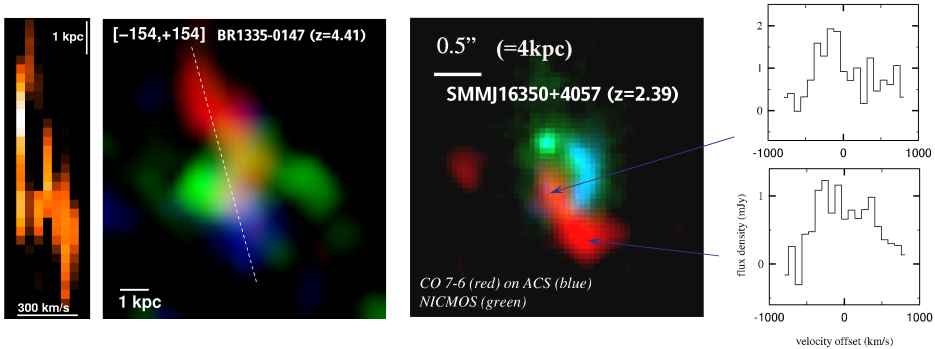}
\caption{{\bf Left:} Position-velocity diagram and composite colour map of the
CO $J=2-1$ emission towards the QSO BRI\,1335$-$0417 ($z=4.41$) at $\sim
0.3\arcsecs$ (FWHM) resolution. CO emission is observed across the velocity
range indicated in the top left corner, with red, green and blue colours
indicating redshifted, central and blueshifted emission relative to the
systemic redshift. Adapted from Riechers et al.\ (2008). {\bf Right:}
Velocity-integrated CO $J=7-6$ emission (shown in red) towards the SMG
SMM\,J16350$+$4057 ($z=2.39$) at $\sim 0.5\arcsecs$ (FWHM) resolution, with the
line profiles from two distinct regions within the source also shown. Blue and
green colours show the rest-frame UV and optical morphologies, respectively.
Adapted from Tacconi et al.\ (2008).}
\end{figure}

Sub-arcsecond interferometric CO observations of SMGs with the IRAM PdBI,
capable of spatially resolving their gas kinematics, have revealed compact
(FWHM diameter $\ls 1.6\,$kpc), sometimes multiple, morphologies with velocity
gradients of $\sim 500\,\rm{km}\,\rm{s}^{-1}$, suggestive of compact rotating
disks (Fig.\ 2). Although the individual components are compact, they are in
some cases found up to $\sim 10\,$kpc apart, yet clearly part of the same
system.  The invoked central gas densities are substantially higher than those of
starforming $z\sim 2$ galaxies selected in the UV/optical, suggesting that SMGs
may represent the end phase of a highly dissipative (major) merger between
gas-rich galaxies (Tacconi et al.\ 2006, 2008).

Other observational highlights have come from high-resolution CO observations
of $z>3$ QSOs (Fig.\ 2), which have successfully probed the gas kinematics on
$\ls 1\,$kpc scales during a period where the host galaxy experiences both
extreme starburst activity and accretion onto its central supermassive black
hole.  When imaged on sub-kpc scales, the molecular gas reservoirs are
often found to break up into multiple, compact disk-like components (e.g.\
Carilli et al.\ 2002; Riechers et al.\ 2008) -- a morphology not too dissimilar
to that of $z\sim 2$ SMGs.  However, the CO line profiles of QSOs are on
average narrower ($\langle\rm{FWHM}\rangle\simeq 300\,\rm{km}\,\rm{s}^{-1}$)
than in SMGs ($\langle\rm{FWHM}\rangle\simeq 700\,\rm{km}\,\rm{s}^{-1}$), and
more often resembles a single Gaussian profile: only $\sim 10\%$ show evidence
of double peaks, yet for SMGs this fraction is $\sim 30-50\%$ (Greve et al.\
2005; Carilli \& Wang 2006).  While a number of factors, such as significantly
different host masses and/or sizes, could affect these findings, it is more
likely that it reflects the (optical) selection of QSO which, unlike the SMG
selection, prefers systems where the gas is rotating in a disk parallel with
the sky plane (Carilli \& Wang 2006).

Whether a picture similar to that seen in $z>3$ QSOs holds for HzRGs is
currently not known. In a few of the most powerful HzRGs there is evidence from
(low-resolution) CO observations that the molecular gas is extended on scales
of $\gs 5\,$kpc, is spatially and kinematically offset from the radio galaxy,
and exhibit velocity-gradients $\gs 500\,\mbox{km}\,\mbox{s}^{-1}$ (e.g.\ De
Breuck et al.\ 2003). These findings are consistent with HzRGs being massive
proto-ellipticals undergoing a major starburst, yet since there has been no
sub-arcsecond CO survey of HzRGs to date, one cannot rule out scenarios in
which the CO traces large-scale molecular outflows, possibly compressed
halo-gas induced by the radio galaxy jets, or streaming, non-virialized gas
caused by close-encounters.

\smallskip

High-resolution CO observations of high-$z$ galaxies also provide a unique
opportunity to couple the dynamical masses of galaxy spheroids
($M_{\mbox{\tiny{sph}}}$, inferred from CO) to the masses of their central
supermassive black holes ($M_{\mbox{\tiny{BH}}}$, inferred from optical
spectroscopy) at epochs where both were undergoing significant growth.
Understanding the physical origin of the locally observed $M_{\mbox{\tiny{BH}}}
- M_{\mbox{\tiny{sph}}}$ relation (Magorrian et al.\ 1998) is a key
scientific goal, as it suggests a fundamental, yet unknown, connection between
black hole accretion and star formation. 

From a comprehensive CO survey of optically luminous ($M_{\mbox{\tiny{B}}}\sim
-28$), submm-detected QSOs at $z\sim 2$, Coppin et al.\ (2008) estimated an
average $M_{\mbox{\tiny{BH}}}/M_{\mbox{\tiny{sph}}}$ ratio of $\sim 9\times
10^{-3}$, nearly an order of magnitude above the local ratio ($\sim
1.4\times 10^{-3}$), while CO observations of $z>4$ QSOs suggest even higher
$M_{\mbox{\tiny{BH}}}/M_{\mbox{\tiny{sph}}}$ ratios (e.g.\ Walter et al.\
2004).  In contrast, CO studies of SMGs (in conjunction with X-ray and
optical/near-IR observations) have shown that SMGs lie a factor of $\sim 3-5$
below the local $M_{\mbox{\tiny{BH}}}-M_{\mbox{\tiny{sph}}}$ relation
(Alexander et al.\ 2008). Clearly, the
$M_{\mbox{\tiny{BH}}} - M_{\mbox{\tiny{sph}}}$ relation evolves with redshift, but
it appears that the evolution occurs differently for different types of galaxies. In
optically luminous QSOs, the build-up of the spheroid stellar mass lags the
the black hole growth, while in SMGs the situation appear to be reversed.  A
level of uncertainty is inherent in these conclusions, however, owing to the
sometimes poorly constrained CO sizes and inclination angles.  More worrying,
though, for such potentially dynamically unsettled systems is the possibility
that the molecular gas might not always be centered around the black hole or
even be a good probe of the dynamical mass. Examples of such AGN/H$_2$
configurations have recently been found at both high and low $z$
(Ivison et al.\ 2008; Riechers et al.\ 2008; Papadopoulos et al.\ 2008).  

\section{The star formation efficiency of galaxies across Cosmic time}
The well known Schmidt/Kennicutt star formation law (Schmidt 1959; Kennicutt
1998), which relate the surface density of star formation rate to the surface
gas density ($\Sigma_{\mbox{\tiny{SFR}}}\propto
\Sigma_{\mbox{\tiny{gas}}}^{1.4\pm 0.2}$), manifests itself in locally observed
relations between IR luminosity (gauging the star formation rate) and molecular
line luminosity (tracing the gas). The best known is the IR-CO relation
($L_{\mbox{\tiny{IR}}}\propto {L'_{\mbox{\tiny{CO}}}}^{1.4-1.7}$), a non-linear
relation when averaged over local spirals, LIRGs and ULIRGs.  In fact the slope
depends on  the particular galaxy sample: it is linear ($N\simeq 1$) for normal
spirals and superlinear ($N\simeq 1.4-1.7$) for starbursts and (U)LIRGs (Gao
2007).
\begin{figure}
\includegraphics[angle=0,width=1.0\hsize]{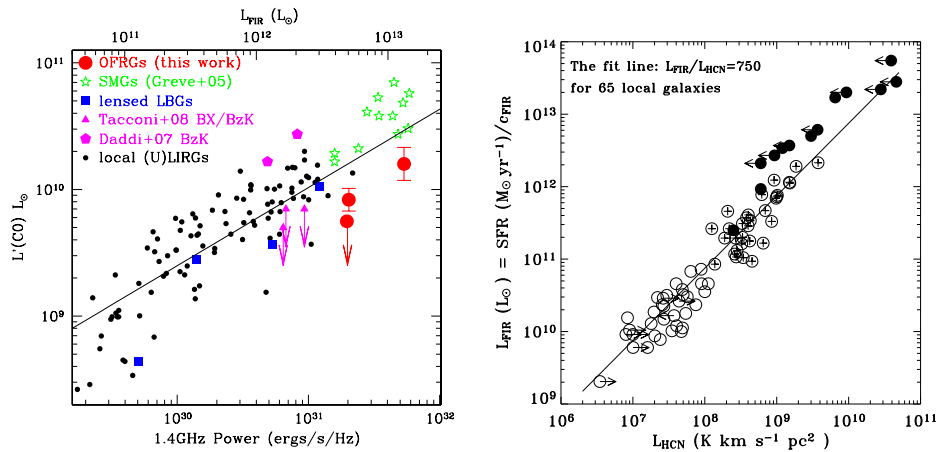}
\caption{{\bf Left:} CO vs.\ radio luminosity for various samples of high-$z$
populations (adapted from Chapman et al.\ 2008).  This may be a cleaner way of
representing the IR-CO correlation since the conversion from submm and/or radio
fluxes to IR-luminosities is notoriously uncertain.  {\bf Right:} The IR-HCN
relation for the 13 $z>1$ sources with HCN observations at the present date
(filled symbols). Adapted from Gao et al.\ (2007).}
\end{figure}

Greve et al.\ (2005) found that bright SMGs follow the non-linear IR-CO
relation established for local (U)LIRGs, thereby extending the relation to the
distant Universe and the highest IR luminosities ($L_{\mbox{\tiny{IR}}}\simeq
10^{13}\,\Lsolar$). Adopting $L_{\mbox{\tiny{IR}}}/{L'_{\mbox{\tiny{CO}}}}$ as
a measure of the star formation efficiency ($SFE \propto SFR/M(\rm{H}_2$), this
would indicate that SMGs form stars more efficiently than local galaxies.
Possible examples of even higher star formation efficiencies are OFRGs (Chapman
et al.\ 2008), which have $\sim 4\times$ higher
$L_{\mbox{\tiny{IR}}}/{L'_{\mbox{\tiny{CO}}}}$ ratios than typical SMGs (Fig.\
3).  Galaxies with low $SFE$s are also encountered at high redshifts, as
demonstrated by Daddi et al.\ (2008) who found
$L_{\mbox{\tiny{IR}}}/{L'_{\mbox{\tiny{CO}}}}$ ratios comparable to that of
local spirals in two $z\sim 1.5$ sBzK galaxies, i.e.\ an order of magnitude
below that of SMGs.

Going instead to a dense gas tracer such as HCN, one finds a linear star
formation law extending from Galactic clouds to spirals and (U)LIRGs (Gao \& Solomon 2004;
Wu et al.\ 2005), indicating that the star formation rate per unit dense gas (as measured
by $L_{\mbox{\tiny{IR}}}/L'_{\mbox{\tiny{HCN}}}$) is virtually independent of
the star formation environment (Fig.\ 3). To date observations of HCN at
high-$z$ have resulted in only five detections and 8 upper limits (Gao et al.\
2007 and references therein), and as a result the nature of the IR-HCN relation
at high-$z$ is uncertain.  There is tentative evidence, however, that the
high-$z$ galaxies, and possibly even the most luminous local ULIRGs depart from
a linear IR-HCN relation (Graci\'{a}-Carpio et al.\ 2008).  Recent theoretical models
tie such $L_{\mbox{\tiny{line}}}-L_{\mbox{\tiny{IR}}}$ relations to a
common underlying H$_2$ cloud density hierarchy and a $SFR \propto \langle
\rho_{\mbox{\tiny{gas}}}\rangle ^{1.5}$ relation (e.g.\ Krumholz \& Thompson
2007; Narayanan et al.\ 2008).  The exponent in the
$L_{\mbox{\tiny{line}}}-L_{\mbox{\tiny{IR}}}$ relation is then a sensitive
function of the $n_{\mbox{\tiny{crit}}}/\langle n\rangle$ ratio, where
$n_{\mbox{\tiny{crit}}}$ is the critical density of the line being observed.
However, the strong variations of HCN $4-3/1-0$ ratios ($\sim 0.1-1$) found
even among ULIRGs with similar IR, HCN, CO $J=1-0$ luminosities (Papadopoulos 2007) 
as well as possible influences of a hard, AGN-originating X-ray spectrum
on the HCN abundance (e.g.\ Lepp \& Dalgarno 1996) may make the tracing of
dense gas far from straightforward.

\section{The physical properties of the ISM in distant galaxies}
A detailed picture of the physical properties of the ISM (e.g.\ abundances,
temperature and density distribution, and thermal balance) will require amble
sampling of the CO, HCN, and HCO$^+$ rotational ladders, along with key
diagnostic atomic transitions such as [C{\sc ii}]\,158\,$\mu$m, [C{\sc
i}]\,369\,$\mu$m, and [C{\sc i}]\,609\,$\mu$m.  Already, multiple CO
transitions have been observed in a number of high-$z$ QSOs and SMGs (Wei\ss~et
al.\ 2007), yielding the first insights into the full CO SLEDs of galaxies at
any redshift (Fig.\ 4). The CO SLEDs of QSOs typically peak at higher
transitions ($J=8-7$) than those of SMGs ($J=6-5$), indicating that the former
harbour denser and warmer gas.
\begin{figure}
\includegraphics[angle=0,width=1.0\hsize]{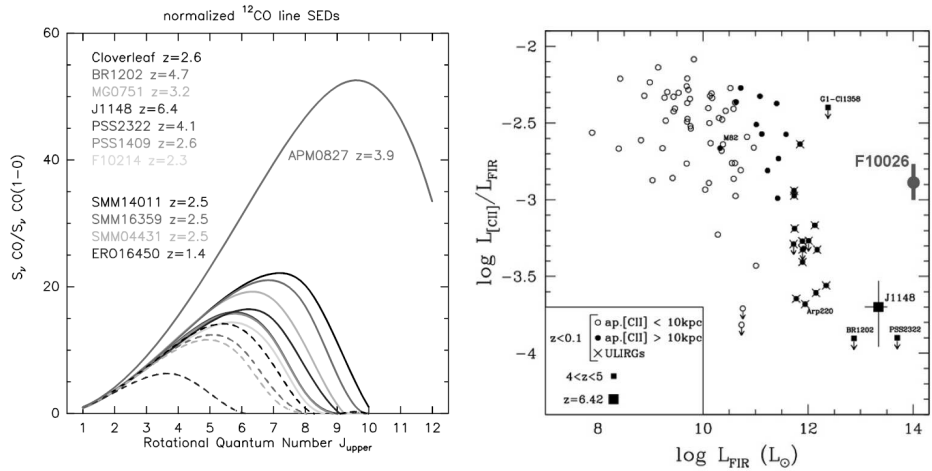}
\caption{{\bf Left:} Modeled CO SLEDs based on multi-line CO survey of 7 QSOs
and 4 SMGs between $z=1.4$ and $z=4.7$. Adapted from Weiss~et al.\ (2007). {\bf
Right:} The $L_{\mbox{\tiny{C{\sc ii}}}}/L_{\mbox{\tiny{IR}}}$ ratios for
normal starforming galaxies, ULIRGs and the only two $z>1$ galaxies with [C{\sc
ii}] detections to date, namely SDSS\,J1148$+$5152 ($z=6.42$) and
FSC\,10026$+$4946 ($z=1.12$).  Upper limits for BR\,1202$-$0725 ($z=4.69$) and
PSS\,J2322$+$1944 ($z=4.11$) are also shown. Adapted from Hailey-Dunsheath et
al.\ (2008).}
\end{figure}

If these observed CO SLEDs are representative of the first objects emerging
from the epoch of reionization at $z>7$, it will have ramifications for the
ability of ALMA to detect CO in such objects. ALMA, with its lack of receivers
below 84\,GHz, will only be able to observe the $J=8-7$ line and above for such
objects. However, these lines are generally not highly excited in $z\sim 2-6$
QSOs, let alone in SMGs, suggesting that such observations will be challenging
even with ALMA (Walter \& Carilli 2008). The situation might even be worse,
since the currently known high-$z$ CO SLEDs are biased towards high-$J$ by a)
strong lensing (in some cases), which favours compact high-$J$ emitting regions
over the more extended low-$J$ emitting gas; b) a general pre-selection of
high-$z$ objects as strong IR/far-IR emitters before CO follow-up is carried
out, which means low excitation CO SLEDs (such as that of HR\,10 --
Papadopoulos \& Ivison 2002) could be missed; c) a further pre-selection in the
sense that high-$z$ objects are first detected in high-$J$ CO lines before
low-$J$ lines are attempted (see \S~2).  This could herald bad news for the use
of high-$J$ lines to yield an unbiased view of the $z > 7$ Universe, since the
CO SLED turn-over may happen at $J=4-3$ for many systems, including the
starbursting ones when between star formation bursts. For these objects, the
EVLA (and possibly the SKA) is likely to be the facilities of choice, since they will
have the frequency coverage and sensitivity to detect the low-$J$ CO lines.

The detection of the [C{\sc ii}] 158\,$\mu$m line towards J1148$+$5251 at
$z=6.42$ (Maiolino et al.\ 2005), the first of its kind, indicated a more
feasible pathway for ALMA to study $z>7$ objects. Not only is the [C{\sc ii}]
line observable with ALMA at these redshifts, but in J1148$+$5251 it is $\sim
5\times$ brighter than the brightest CO line observed in this source ($J=6-5$),
suggesting that [C{\sc ii}] 158\,$\mu$m will be the line of choice when
studying the earliest objects with ALMA, and that a strong synergy will develop 
between ALMA ([C{\sc ii}]) and the EVLA (low-$J$ CO lines).

The [C{\sc i}] and [C{\sc ii}] lines have great diagnostic power in determining
the physical properties of the ISM. Neutral carbon is well mixed in cloud
interiors and holds the promise of probing the bulk H$_2$ gas mass and its
temperature (via the [C{\sc i}] 2-1/1-0 ratio) at high redshifts without being
tied to assumptions about virialized clouds. Testing its H$_2$-tracing
capabilities in local starbursts have yielded results in accordance with the
standard CO-based method (Gerin \& Phillips; Papadopoulos \& Greve 2004).  The
[C{\sc ii}] 158\,$\mu$m line is the main cooling agent of the ISM in our own
Galaxy and in typical starbursts in the local Universe, where it carries a
significant fraction of the IR total luminosity ($L_{\mbox{\tiny{C{\sc
ii}}}}/L_{\mbox{\tiny{IR}}} \simeq 5-10\times 10^{-3}$).  However, cooling via
[C{\sc ii}] is about an order of magnitude less efficient in local ULIRGs
($L_{\mbox{\tiny{C{\sc ii}}}}/L_{\mbox{\tiny{IR}}} \simeq 2-5\times 10^{-4}$ --
e.g.\ Gerin \& Phillips 2000) than in normal starburst galaxies. This is a
trend followed by J1148$+$5152 (Fig.\ 4), and raises the question of how the
gas cools in very luminous galaxies. Recent high-$J$ CO observations of the
local IR-luminous galaxy Mrk\,231 indicate that the CO emission from the dense
gas may produce about the same amount of cooling radiation as the [C{\sc ii}]
line (Papadopoulos et al.\ 2007), suggesting a totally different thermal
balance than in lower luminosity galaxies. On the other hand, [C{\sc ii}] would
be more ubiquitous than high-$J$ CO lines in the ISM of galaxies, across a wide
swath of conditions: not only does it remain luminous in metal-poor systems
(e.g.\ Lyman-alpha emitters) but it is also not tied to star-forming sites; in
fact even the CNM and WNM H{\sc i} phases have significant [C{\sc ii}] emission
contributions.  The latter makes this line an excellent dynamical mass probe,
but also though to interpret in terms of H$_2$ mass alone since non-trivial
corrections must be made for [C{\sc ii}] emission from atomic and even ionized
gas (e.g.\ Madden et al.\ 1997).

\section{What will the future bring?}
The combined efforts of future cm/(sub)mm facilities (ALMA, CCAT, LMT), and
upgraded existing facilities (IRAM PdBI+30m, CARMA, eSMA, EVLA, GBT, e-Merlin),
will allow for unbiased, large-scale surveys of molecular and atomic lines in
galaxies out to $z\sim 10$. One could envisage multiplexed, broadband cm/mm
spectrometers (akin to Z-Spec and ZEUS) on large single-dish (sub)mm telescopes
such as CCAT and LMT carry out large 'blind' molecular line searches that
objectively sample line luminosity functions (in terms of redshift and
luminosity). ELVA and ALMA would facilitate detailed studies via the low- to
mid-$J$ CO/HCO$^+$/HCN lines and atomic transitions ([C{\sc i}]/[C{\sc ii}] and
[N{\sc ii}]).  Herschel will observe the high-$J$ ($J > 5-4$) CO lines
in local (U)LIRGs, resulting in fully sampled CO SLEDs in the local Universe,
which undoubtedly will be of great use in our exploration of distant galaxies.
In the next decade, therefore, we expect to see a genuine revolution in our
ability to study the cosmic star formation and ISM evolution, resulting in
definite answers to all (and many more) of the issues outlined in this text.

\bigskip

\noindent{\bf Acknowledgment} I thank the conference organizers for a
wonderful meeting. I'm also grateful to my collaborators P.\ P.\ Papadopoulos, R.\ J.\ Ivison,
A.\ W.\ Blain, L.\ Hainline, I.\ Smail, F.\ Bertoldi, S.\ C.\ Chapman, R.\
Genzel, L.\ Tacconi, R.\ Neri.


\end{document}